\documentclass[twocolumn]{article}
\usepackage{authblk}

\usepackage{array,booktabs} 
\usepackage{amsmath,amssymb} 
\usepackage{stfloats} 
\usepackage{natbib}
\usepackage{hyperref}
\usepackage{color}
\usepackage{xcolor}
\usepackage[utf8]{inputenc}
\usepackage{flushend}
\usepackage{newtxtext,newtxmath}
\definecolor{darkgreen}{rgb}{0,0.75,0}
\usepackage{glossaries}
\usepackage{silence}
\makeglossaries
\usepackage[per-mode = symbol]{siunitx}
\usepackage{tabularx}
\usepackage{todonotes}

\hypersetup{
	pdfauthor={Bjørn-Olav H. Eriksen, Morten Breivik},
	pdftitle={Short-term ASV Collision Avoidance with Static and Moving Obstacles},
	pdfkeywords={Autonomous surface vehicles, collision avoidance, model predictive control (MPC)},
	colorlinks=true,
	citecolor=darkgreen,
	urlcolor=darkgreen,
	pdfstartview=Fit,
	pdfpagelayout=SinglePage,
	pdfcreator=pdflatex,
	pdfproducer=pdflatex}

\newacronym{asv}{ASV}{autonomous surface vehicle}
\newacronym{auv}{AUV}{autonomous underwater vehicle}
\newacronym{ais}{AIS}{automatic identification system}
\newacronym{colav}{COLAV}{collision avoidance}
\newacronym{colregs}{COLREGs}{the International Regulations for Preventing Collisions at Sea}
\newacronym{vo}{VO}{velocity obstacle}
\newacronym{dw}{DW}{dynamic window}
\newacronym{a*}{A*}{A star}
\newacronym{rrt}{RRT}{rapidly exploring random tree}
\newacronym{bcmpc}{BC-MPC}{branching-course model predictive control}
\newacronym{sog}{SOG}{speed over ground}
\newacronym{rot}{ROT}{rate of turn}
\newacronym{mr}{MR}{Maritime Robotics}
\newacronym{fffb}{FF-FB}{feedforward feedback}
\newacronym{pdaf}{PDAF}{probabilistic data association filter}
\newacronym{ros}{ROS}{Robot Operating System}
\newacronym{los}{LOS}{line of sight}
\newacronym{dof}{DOF}{\:degrees of freedom}
\newacronym{pi}{PI}{proportional-integral}
\newacronym{fffbc}{FF-FB-C}{feedforward-feedback course}
\newacronym{fbc}{FB-C}{feedback course}
\newacronym{iae}{IAE}{integral of absolute error}
\newacronym{iadc}{IADC}{integral of absolute differentiated control}
\newacronym{iae-adc}{IAE-ADC}{integral of absolute error times the integral of absolute differentiated control}
\newacronym{iaew}{IAEW}{integral of absolute error times work}
\newacronym{mpc}{MPC}{model predictive control}
\newacronym{osd1}{OSD1}{Ocean Space Drone 1}
\glsdisablehyper
\begin{document}
\newcommand{\skewSymThree}[3]{\begin{bmatrix}0 & -#3 & #2 \\ #3 & 0 & -#1 \\ -#2 & #1 & 0\end{bmatrix}}
\newcommand{\bs}{\boldsymbol}
\newcommand{\norm}[1]{\left\lVert#1\right\rVert}
\newcommand{\tr}{^\top}

\title{Short-term ASV Collision Avoidance \\with Static and Moving Obstacles}


\author{Bjørn-Olav H. Eriksen}
\author{Morten Breivik}

\affil{Centre for Autonomous Marine Operations and Systems, Department of Engineering Cybernetics, Norwegian University of Science and Technology (NTNU), NO-7491 Trondheim, Norway. E-mail: \textsf{\{bjorn-olav.h.eriksen, morten.breivik\}@ieee.org}}

\maketitle

\begin{abstract}
This article considers \gls{colav} for both static and moving obstacles using the \gls{bcmpc} algorithm, which is designed for use by \glspl{asv}.
The \gls{bcmpc} algorithm originally only considered \gls{colav} of moving obstacles, so in order to make the algorithm also be able to avoid static obstacles, we introduce an extra term in the objective function based on an occupancy grid.
In addition, other improvements are made to the algorithm resulting in trajectories with less wobbling.
The modified algorithm is verified through full-scale experiments in the Trondheimsfjord in Norway with both virtual static obstacles and a physical moving obstacle.
A radar-based tracking system is used to detect and track the moving obstacle, which enables the algorithm to avoid obstacles without depending on vessel-to-vessel communication.
The experiments show that the algorithm is able to simultaneously avoid both static and moving obstacles, while providing clear and readily observable maneuvers.
The \gls{bcmpc} algorithm is compliant with rules 8, 13 and 17 of the \gls{colregs}, and favors maneuvers following rules 14 and 15.
\end{abstract}

\providecommand{\keywords}[1]{\textbf{\textit{Keywords: }} #1}

\section{Introduction}\label{sec:Introduction}
All parts of society are currently being automated at a rapid pace.
One example is the development of autonomous cars, as exemplified by the development efforts made by e.g. Tesla, Google and Uber.
Such a trend is also ongoing in the maritime domain, where autonomous technology presents opportunities for increased cost efficiency, in addition to reducing the environmental impact of goods and passenger transport.
One example of this is the Yara Birkeland project in Norway, where an electrically-powered autonomous cargo ship will replace $40000$ diesel-powered truckloads of fertilizer each year by 2022~\citep{Paris2017}.
Furthermore, it is reported that in excess of 75\% of maritime accidents are caused by human errors \citep{Chauvin2011,Levander2017}, which also reveals a potential for increased safety by introducing autonomous technology at sea.
Employing \glspl{asv} in areas where other vessels are present does, however, require a robust \gls{colav} system in order to avoid collisions and operate safely.

\begin{figure}
	\centering
	\includegraphics[width=\linewidth]{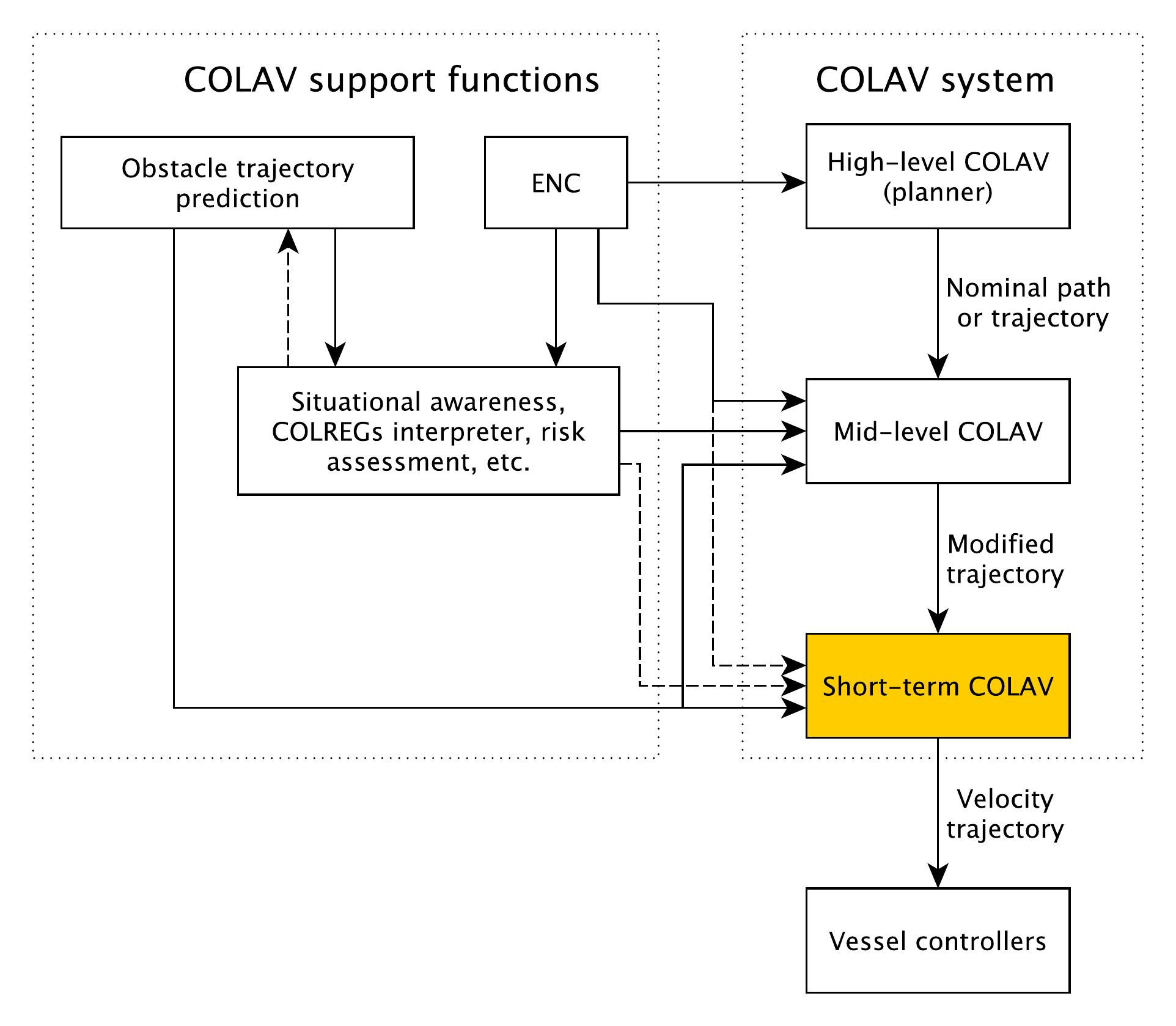}
	\caption{\label{fig:hybrid}A hybrid architecture with three layers.
	The support functions provide relevant information for the COLAV algorithms, including prediction of obstacle trajectories, static obstacles from electronic nautical charts (ENC) and situational awareness in the form of COLREGs situations.}
\end{figure}

There exists several algorithms for \gls{asv} \gls{colav}, e.g. the \gls{vo} algorithm \citep{Kuwata2014}, the A* algorithm \citep{Schuster2014} and algorithms based on \gls{mpc} and optimization \citep{Benjamin2006,Svec2013,Abdelaal2016,Hagen2018}.
These algorithms are, however, designed with the idea of ``one size fits all'', where the same algorithm is used to solve both situations requiring proactive and reactive behaviors.
A challenge with this approach is that the algorithm must be able to solve problems of a wide range sufficiently well, which makes the algorithm difficult to design and tune.
A different approach is to utilize a hybrid architecture~\citep{Loe2008,Casalino2009}, where the complementary strengths of different algorithms can be combined in a layered architecture.
An example of a hybrid architecture is shown in Figure~\ref{fig:hybrid}, where the \gls{colav} system is divided into three layers, namely a high-level, mid-level and a short-term \gls{colav} algorithm. 
The high-level planner performs long-term planning by finding a path or trajectory from an initial position to a goal position while being able to avoid static obstacles, satisfy time constraints and minimize energy consumption.
The mid-level algorithm attempts to follow the planned path or trajectory from the high-level planner, while making local modifications in order to avoid moving obstacles.
This algorithm should be designed to comply with the maneuvering rules of the \gls{colregs}, which dictates how vessels should behave in situations where there exists a risk of collision with other vessels~\citep{Cockcroft2004}.
The short-term \gls{colav} algorithm inputs the modified trajectory from the mid-level algorithm, and should have low computational requirements ensuring that the \gls{colav} system can react to sudden changes in the environment.
This algorithm should also serve as a final safety barrier in situations where e.g. the mid-level algorithm fails to find a solution~\citep{Eriksen2017b}.
In addition, the short-term algorithm should have a shorter planning horizon than the mid-level algorithm, making it inherently capable of handling situations where the \gls{colregs} may require ignoring the maneuvering aspects of rules 14 and 15 when moving obstacles do not comply with the \gls{colregs}.
The algorithm should, however, maneuver in accordance with rules 14 and 15 when the situation allows it.

The authors have performed a significant amount of work on the hybrid architecture in Figure~\ref{fig:hybrid}, concerning e.g. model-based vessel controllers \citep{Eriksen2017,Eriksen2018b}, short-term \gls{colav} \citep{Eriksen2018,Eriksen2019}, mid-level \gls{colav} \citep{Eriksen2017b} and a high-level planner interfaced to the mid-level algorithm \citep{Bitar2019}.
In an upcoming article \citep{Eriksen2019c}, we populate the hybrid architecture with algorithms including the \gls{bcmpc} algorithm discussed in this article, and demonstrate \gls{colav} compliant with \gls{colregs} rules 8 and 13--17 in simulations.
Work has also been performed on obstacle trajectory prediction \citep{Hexeberg2017,Dalsnes2018}.
For the short-term \gls{colav} layer, we initially focused on the \gls{dw} algorithm, using a radar-based tracking system for detecting and tracking obstacles \citep{Wilthil2017}.
The reason for using exteroceptive sensors such as radars for detecting obstacles is that they do not depend on vessel-to-vessel communication or collaboration with other vessels, hence enabling avoidance of vessels which do not have or use \gls{ais} transponders.
Another questionable aspect of \gls{ais} is that other vessels may provide incorrect information \citep{Harati-Mokhtari2007}, which can be difficult to detect and handle.
However, there is a fair amount of noise on obstacle estimates originating from systems using exteroceptive sensors, which the \gls{dw} algorithm was shown not to handle sufficiently well in full-scale experiments \citep{Eriksen2018}.
We therefore developed the \gls{bcmpc} algorithm for short-term \gls{colav} \citep{Eriksen2019}, which is based on \gls{mpc} and designed to be robust to obstacle estimate noise.
This algorithm is shown to have good performance in full-scale experiments, but originally only accounts for moving obstacles.

In this article, we further develop the \gls{bcmpc} algorithm to also handle avoidance of static obstacles in addition to moving obstacles, as well as producing trajectories with less wobbling.
The modified algorithm is verified in full-scale experiments in Trondheimsfjorden, Norway, showing good performance.
The experiments are performed with virtual static obstacles, while a moving obstacle is detected and tracked using a radar, not depending on vessel-to-vessel communication.

The rest of this article is organized as follows: Section~\ref{sec:the_bc_mpc_algorithm} presents the \gls{bcmpc} algorithm and the modifications we do to it, Section~\ref{sec:experimental_results} presents the experimental setup and results, while Section~\ref{sec:Conclusion} concludes the article and points to possibilities for further work.

\section{The BC-MPC algorithm} 
\label{sec:the_bc_mpc_algorithm}
The \gls{bcmpc} algorithm \citep{Eriksen2019} is a \gls{colav} algorithm designed using sample-based \gls{mpc}, intended for short-term \gls{colav} for \glspl{asv}.
Sample-based \gls{mpc} algorithms are based on computing an objective function over a finite discrete search space and selecting the optimized solution, rather than utilizing search algorithms as in gradient-based algorithms.
A benefit of sample-based algorithms is that they do not have problems with solving highly nonlinear and non-convex problems, which in general is difficult for gradient-based algorithms.
This makes sample-based algorithms well suited for use in the short-term layer in Figure~\ref{fig:hybrid}.  
Furthermore, the \gls{bcmpc} algorithm is designed to be robust with respect to noisy obstacle estimates, which is a significant source of disturbance when using exteroceptive sensors such as radars for detecting and tracking obstacles.

With respect to the \gls{colregs}, the \gls{bcmpc} algorithm complies with rules 8, 13 and 17, and favors maneuvers following rules 14 and 15.
In cases where the algorithm chooses to ignore the maneuvering aspects of rules 14 and 15, which can be required when rule 17 revokes a stand-on obligation, the maneuvers have an extended clearance to obstacles.

At each iteration, the algorithm computes a search space consisting of a finite number of possible trajectories, which each contains a sequence of maneuvers.
Given this search space, an objective function is computed on the trajectories, and the optimized trajectory is selected and used as the reference to the vessel controllers which control the \gls{sog} and course.
The algorithm is based on \gls{mpc}, hence only the first part of the optimized trajectory is used before a new solution is computed and implemented.

This section presents an overview of the \gls{bcmpc} algorithm.
Interested readers are referred to \citet{Eriksen2019} for more details on the algorithm.
In addition, this section presents modifications enabling the algorithm to perform static obstacle avoidance and produce trajectories with less wobbling than the original algorithm.

\subsection{Trajectory generation}
At each iteration, a new finite search space of possible trajectories is generated.
Every trajectory contains a number of sub-trajectories, each containing one maneuver.
This naturally forms a tree structure, with the nodes representing vessel configurations and edges representing maneuvers.
The initial condition is used as the root node, and the depth of the tree is equal to the number of maneuvers in each trajectory.

\begin{figure}[t]
  	\centering
  \begin{tabular}[b]{@{}c@{}}
      \includegraphics[width=\linewidth]{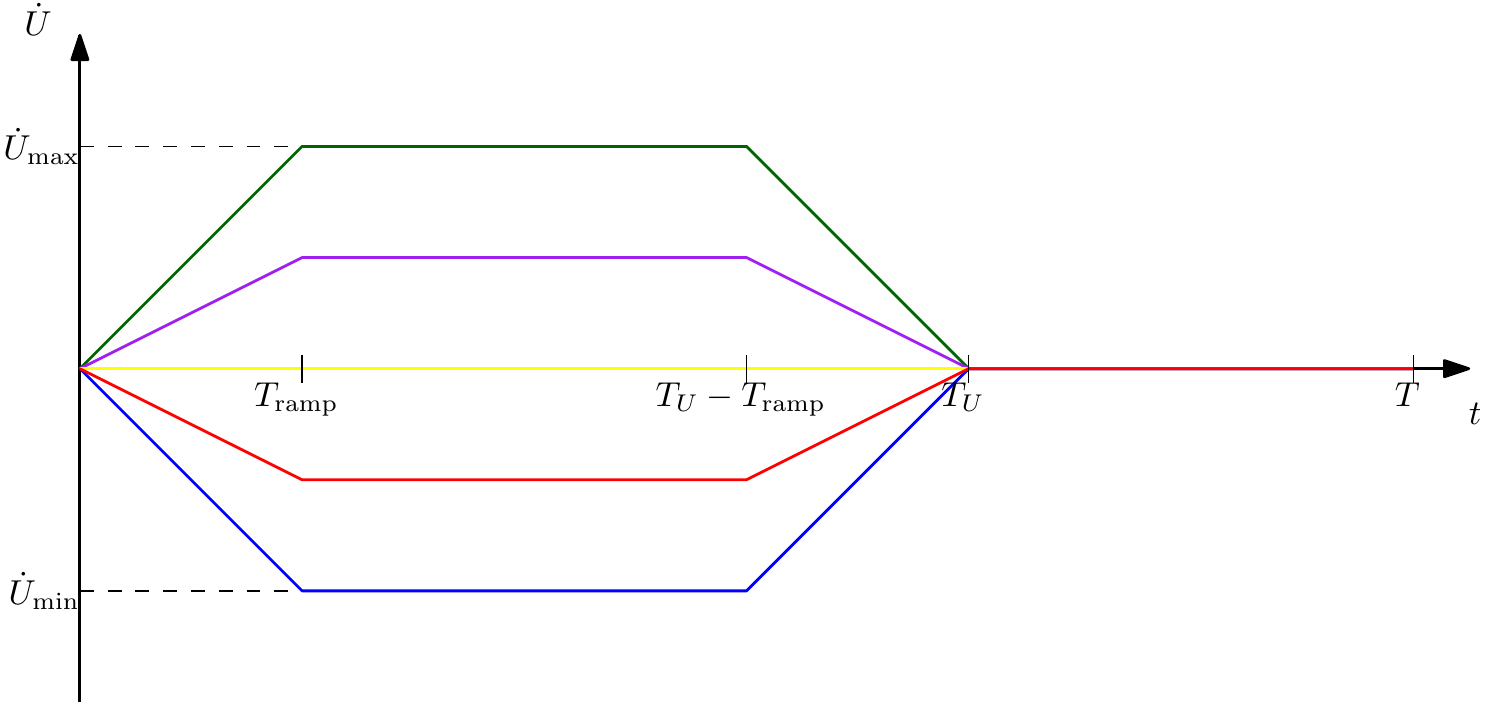} \\
      \small (a) Speed acceleration motion primitives \\
      \includegraphics[width=\linewidth]{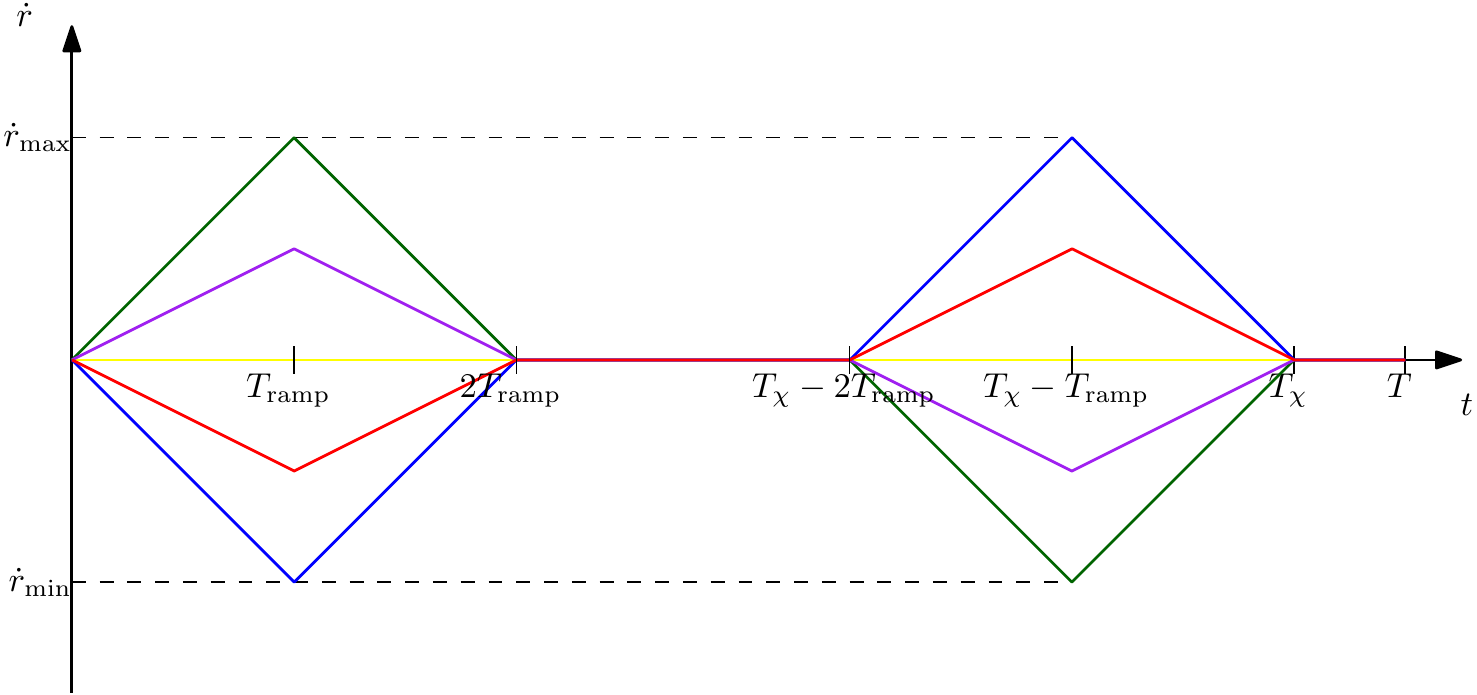} \\
      \small (b) Course acceleration motion primitives.
  \end{tabular}
  	\caption{\label{fig:motionPrimitives}Acceleration motion primitives, where $T$ is the step time, $T_{ramp}$ denotes the ramp time, while $T_U$ and $T_\chi$ are the \gls{sog} and course maneuver time lengths, respectively. The symbols $\dot U_{\max}, \dot U_{\min}, \dot r_{\max}$ and $\dot r_{\min}$ denote the acceleration limits of the vessel at the initial vessel state.}
\end{figure}

The trajectory generation is performed by a repeatable maneuver-generation procedure, which when given a vessel configuration computes a set of sub-trajectories each containing one maneuver.
Piecewise linear acceleration profiles in speed and course serve as a template for the maneuvers.
An example of $5$ motion primitives based on the acceleration profiles in speed and course is shown in Figure~\ref{fig:motionPrimitives}.
The acceleration profiles are dependent on the step time length (the maneuver time length) $T>0$, the ramp time $T_{\text{ramp}}\in(0,\text{min}(\frac{T_U}{2},\frac{T_\chi}{4})]$ and the speed and course maneuver lengths, $T_U, T_\chi \in (0,T]$, respectively.
Given a current vessel velocity, the maximum and minimum speed and course accelerations $\dot U_{\max},\dot U_{\min},\dot r_{\max}$ and $\dot r_{\min}$ are computed using a vessel model.

To improve the convergence properties of the algorithm, we employ a guidance function which can modify some of the trajectories in the search space.
This is done by moving the closest acceleration sample in speed and course to a desired acceleration generated by the guidance function, if this is inside the feasible acceleration region.

\begin{figure}
  \centering
  \includegraphics[width=\linewidth]{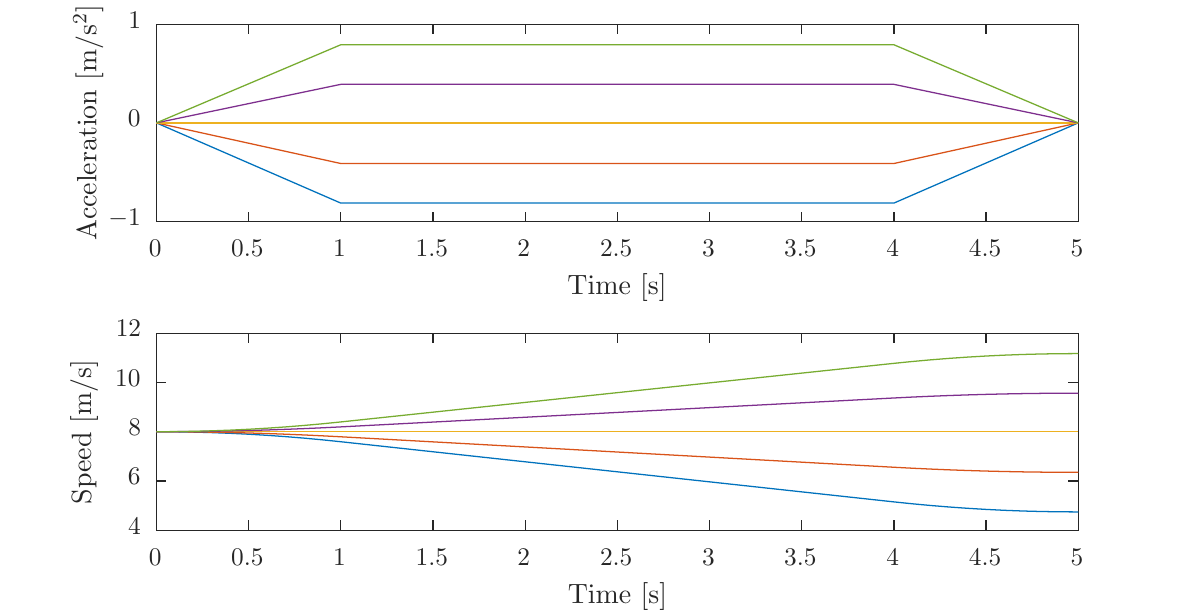}
  \caption{\label{fig:speedtraj}Example of 5 speed trajectories with ramp time $T_{\text{ramp}}=1~\si{\second}$, and maneuver and step time lengths $T_U=T=5~\si{\second}$.
  Acceleration is shown in the top plot, while speed is shown in the bottom plot.}
\end{figure}

\begin{figure}[t]
  \centering
  \includegraphics[width=\linewidth]{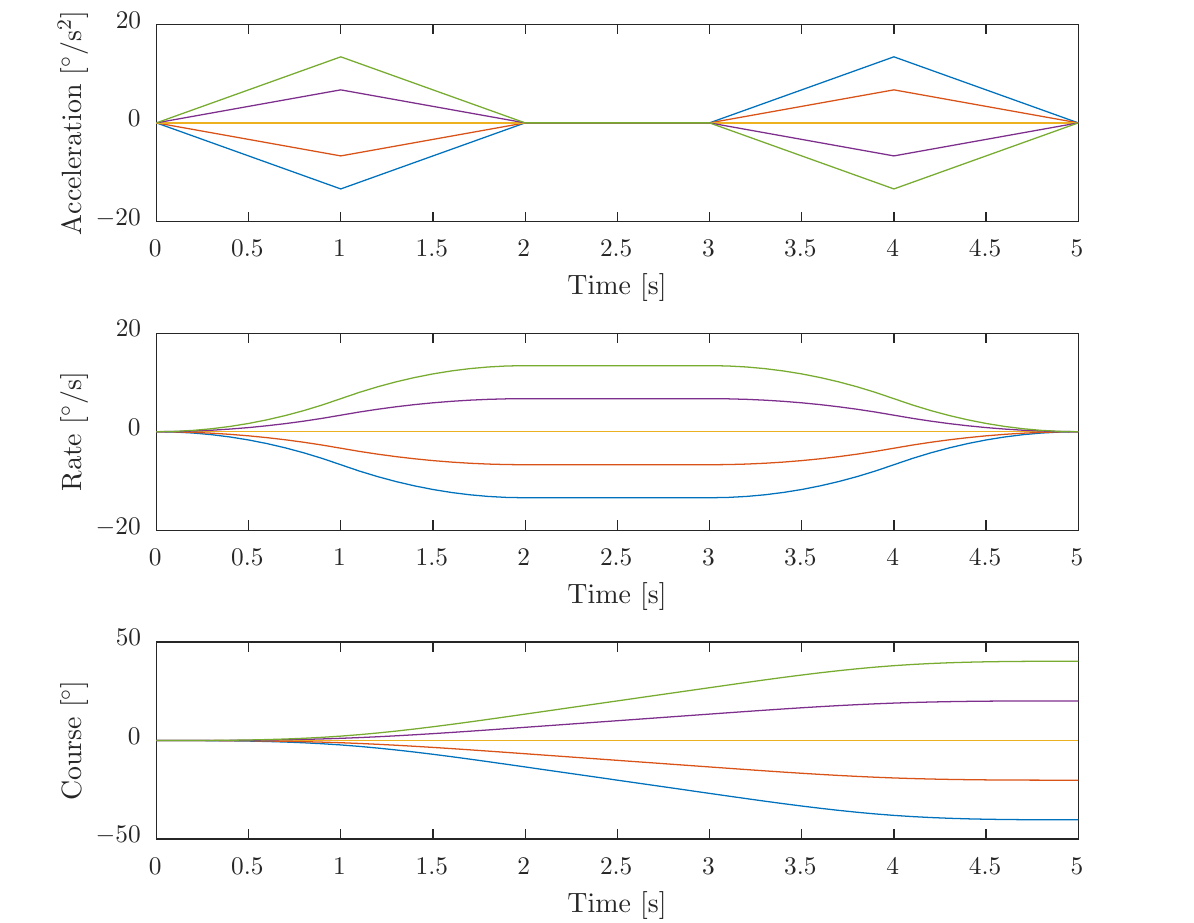}
  \caption{\label{fig:coursetraj}Example of 5 course trajectories with ramp time $T_{\text{ramp}}=1~\si{\second}$, and maneuver and step time lengths $T_\chi=T=5~\si{\second}$.
  Acceleration is shown in the top plot, rate in the middle plot and course in the bottom plot.}
\end{figure}

Desired speed and course trajectories $U_{d}(t)$ and $\chi_d(t)$ are generated by analytically integrating the acceleration motion primitives.
Numerical examples of 5 speed and 5 course trajectories are shown in figures~\ref{fig:speedtraj} and \ref{fig:coursetraj}.
It should be noted that these trajectories are intended as reference trajectories for the vessel controllers, hence they are initiated in an open-loop fashion with the current desired speed and course in order to ensure continuous references for the vessel controllers.
The desired speed and course trajectories are joined together in a union set of desired velocity trajectories:
\begin{multline}\label{eq:desiredVelocities}
	\mathcal{U}_d = \{U_{d,1}(t), U_{d,2}(t), \ldots, U_{d,{N_U}}(t)\} \\\times \{ \chi_{d,1}(t), \chi_{d,2}(t), \ldots , \chi_{d,{N_\chi}}(t)\},
\end{multline}
resulting in a total of $N_U\!\cdot\! N_\chi$ desired velocity trajectories where $N_U \in \mathbb{Z}^+$ and $N_\chi\in \mathbb{Z}^+$ are the number of speed and course motion primitives.
To include feedback in the trajectory generation, we use an error model of the vessel to generate feedback-corrected speed and course trajectories $\bar U_{d}(t)$ and $\bar \chi_d(t)$, which similarly as in \eqref{eq:desiredVelocities} is combined in a set $\bar{\mathcal{U}}_d$.
The feedback-corrected speed and course trajectories are used to generate feedback-corrected predicted pose trajectories:
\begin{equation}\label{eq:predictedPosition}
	\bar{\mathcal{H}} = \left\{\bar {\bs\eta}(t; \bar U(t), \bar \chi(t)) \big | (\bar U(t), \bar \chi(t)) \in \bar{\mathcal{U}} \right\},
\end{equation}
where $\bar {\bs\eta}(t; \bar U(t), \bar \chi(t))$ denotes a kinematic simulation procedure to obtain the vessel pose.

\begin{figure}[t]
  \centering
  \includegraphics[width=\linewidth]{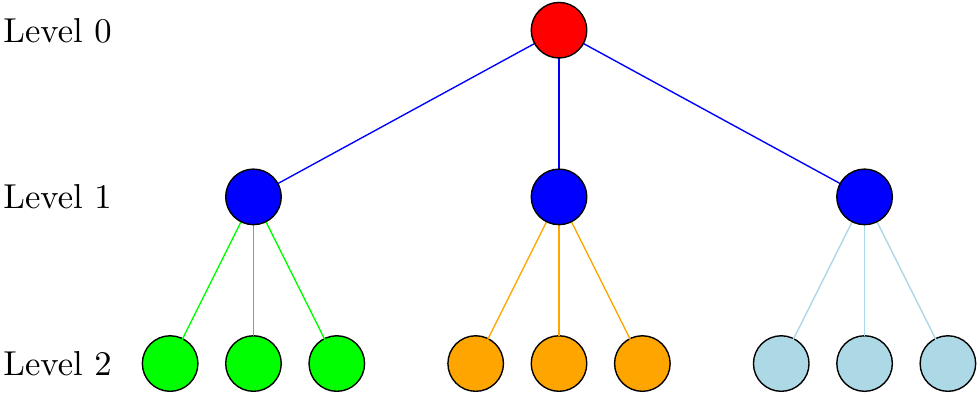}
  \caption{\label{fig:trajTree}Illustration of a trajectory prediction tree with two levels.
  The red node is the root node containing the initial vessel configuration.
  Other colors group nodes and edges associated with each maneuver-generation procedure, which generate three maneuvers each time (given by combinations of $N_U$ and $N_\chi$ satisfying $N_U\!\cdot\! N_\chi = 3$).
  The tree contains a total of nine trajectories, each consisting of two sub-trajectories.
  }
\end{figure}

A full trajectory search space is created by first generating a set of sub-trajectories by using the maneuver-generation procedure initialized with the initial vehicle pose.
At this stage, the prediction tree has a depth of one with the initial vessel pose as the root node and a set of leaf nodes each reached by one maneuver.
Following this, we append the trajectories with another maneuver by repeating the maneuver-generation procedure, initialized on each of the leaf nodes, which increases the depth of the trajectory prediction tree with one level.
This is repeated until the trajectory prediction tree has the desired depth, i.e. each trajectory has the desired number of maneuvers.
This concept is illustrated in Figure~\ref{fig:trajTree}.
The acceleration profile parameters and number of speed and course motion primitives can be level-dependent, which allows for shaping the maneuvers differently and avoiding exponential growth with the number of levels.
To reduce the complexity in tuning the algorithm, we use the same ramp time $T_{\text{ramp}}$ and speed and course maneuver lengths $T_U$ and $T_\chi$ throughout each level.
For a desired trajectory tree depth $B$ ($B$ maneuvers in each trajectory), this leaves us with deciding the step time lengths of each level $\bs T = [T_1, T_2, \ldots, T_B]$, and the number of speed and course maneuvers at each level $\bs N_U = [N_{U,1}, N_{U,2}, \ldots, N_{U,B}]$ and $\bs N_\chi = [N_{\chi,1}, N_{\chi,2}, \ldots, N_{\chi,B}]$.

\begin{figure}[t]
	\centering
	\includegraphics[width=\linewidth]{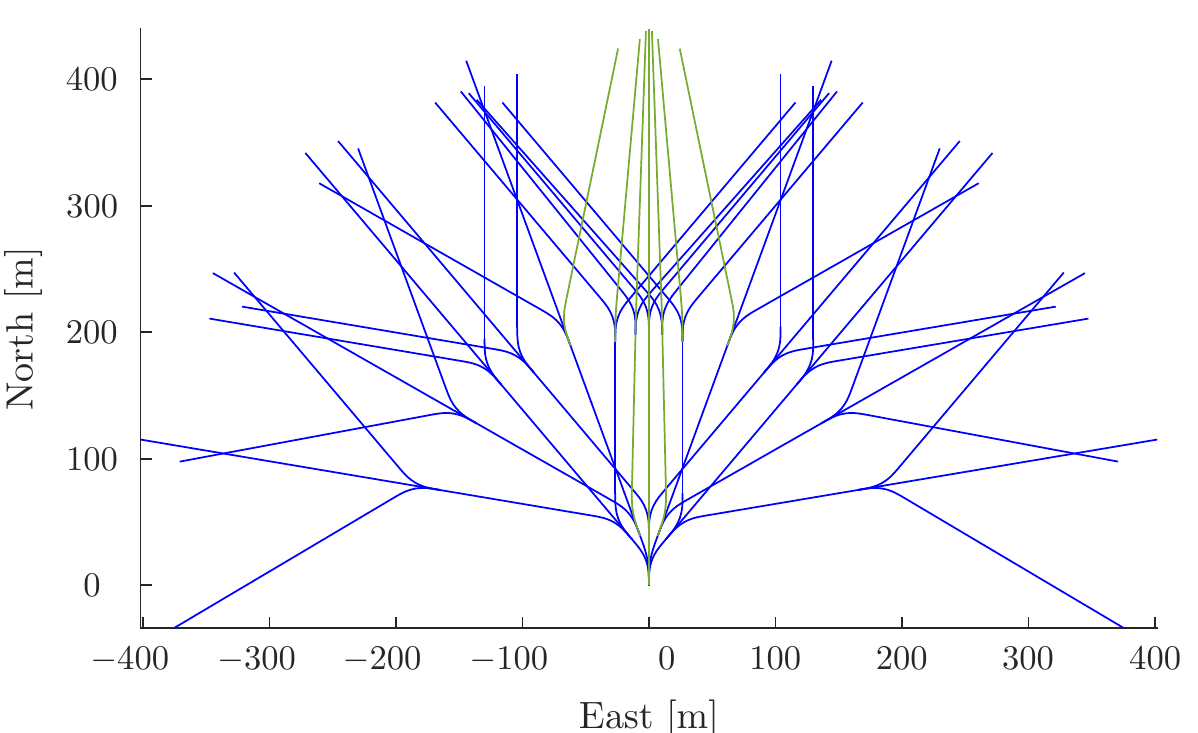}
  	\caption{A set of predicted pose trajectories with three levels. Notice how the guidance function shifts some of the maneuvers, marked in dark green, to converge towards the desired trajectory, which is a straight-north trajectory from the initial pose (not shown in the figure). 
    For illustration purposes, the trajectories only contain course maneuvers.\label{fig:postraj}}
\end{figure}

A set of feedback-corrected predicted pose trajectories for a trajectory generation with $B=3$ levels is shown in Figure~\ref{fig:postraj}.
The ramp time is $T_{\text{ramp}}=1~\si{\second}$, and the speed and course maneuver lengths are $T_U = T_\chi = 5~\si{\second}$.
The step time lengths are $\bs T = [20, 30, 30]~\si{\second}$, and the number of speed and course maneuvers are $\bs N_U = [1,1,1]$ and $\bs N_\chi = [5,3,3]$.

\subsection{Selecting the optimized trajectory}
Given a search space of vessel trajectories and a desired trajectory $\bs p_d(t)\in \mathbb{R}^2$, we solve an optimization problem to find the optimized desired velocity trajectory $\bs u_d^*(t) = \begin{bmatrix} U_d^*(t) & \chi_d^*(t) \end{bmatrix}\tr$ as:
\begin{equation}\label{eq:ocp}
	\bs u_d^*(t) = \underset{(\bar{\bs \eta}_k(t), \bs u_{d,k}(t)) \in (\bar{\mathcal{H}}, \mathcal{U}_d)}{\text{argmin}} G(\bar{\bs \eta}_k(t), \bs u_{d,k}(t); \bs p_d(t)).
\end{equation}
The objective function is given as:
\begin{multline}\label{eq:objFunc}
 	G(\bar{\bs \eta}(t), \bs u_d(t);\bs p_d(t)) = w_{\text{al}}\text{align}(\bar{\bs \eta}(t); \bs p_d(t)) \\+ w_{\text{av,m}}\text{avoid}_\text{m}(\bar{\bs \eta}(t)) + w_{\text{av,s}}\text{avoid}_\text{s}(\bar{\bs \eta}(t)) \\ + w_{\text{t},U}\text{tran}_U(\bs u_d(t)) + w_{\text{t},\chi}\text{tran}_\chi(\bs u_d(t)),
\end{multline}
where $w_{\text{al}},w_{\text{av,m}},w_{\text{av,s}},w_{\text{t},U},w_{\text{t},\chi}>0$ are tuning parameters. 

\begin{figure}[t]
  \centering
  \includegraphics[width=.8\linewidth]{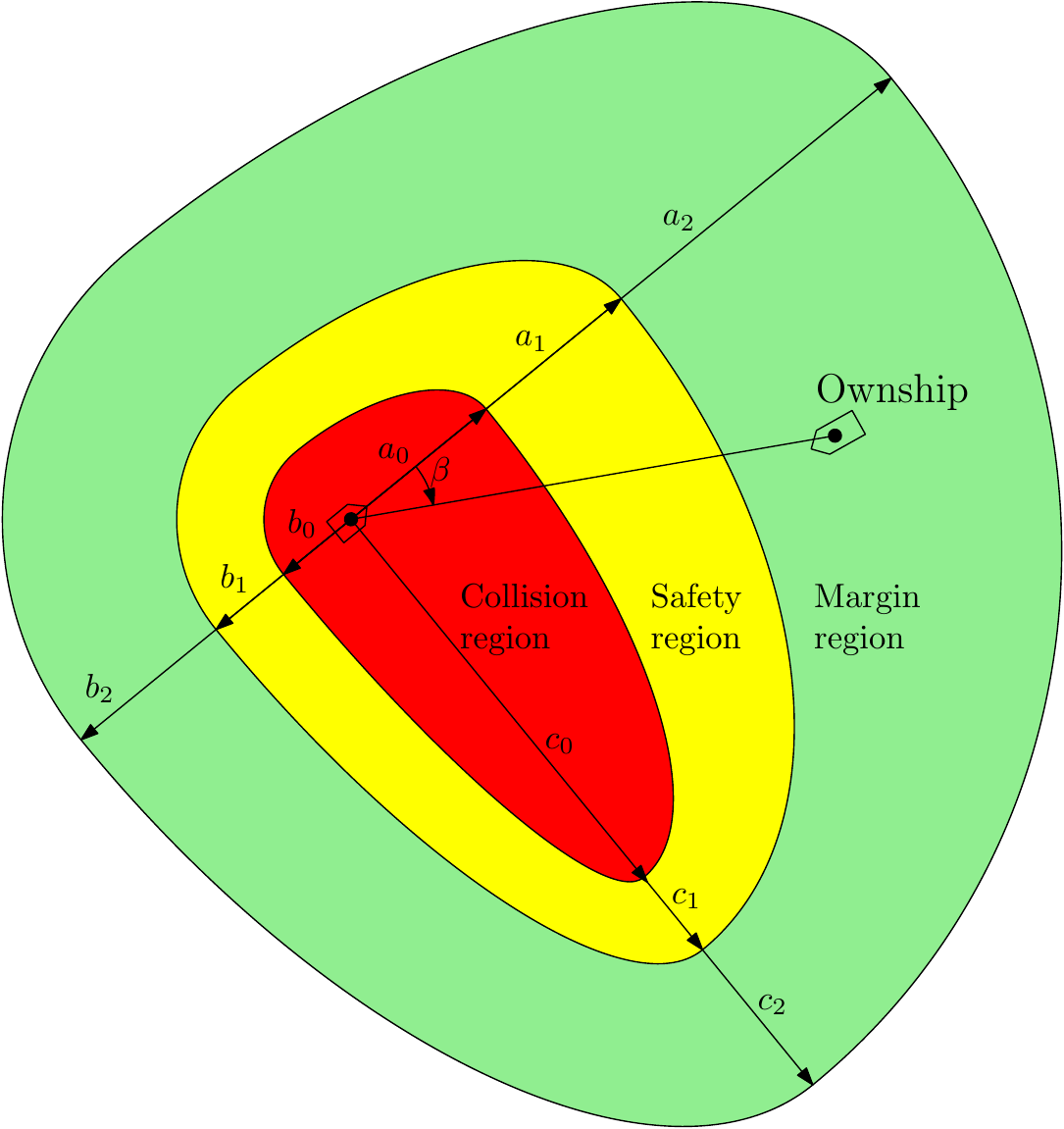}
  \caption{\label{fig:ellipticalColregsObsCostRegions}Avoidance cost regions centered at the moving obstacle, each constructed by one circular and three elliptical segments. The green, yellow and red regions are named the margin, safety and collision regions, respectively. The avoidance cost increases linearly with different gradients inside the green and yellow regions, while the cost is constant inside the red region. The variables $a_i,b_i$ and $c_i$, $i\in\{1,2,3\}$ denote the region sizes, where $c_i = b_i + d_{\text{COLREGs}}$ with $d_{\text{COLREGs}}$ controlling the COLREGs expansion.}
\end{figure}

The $\text{align}(\cdot)$ function assigns a value to following the desired trajectory $\bs p_d(t)$. 
The $\text{avoid}_\text{m}(\cdot)$ function assigns a cost to traveling close to moving obstacles, which depends on the distance to an obstacle for each point on the predicted trajectories.
The maneuvering rules in the \gls{colregs}, rules 13--15, require the vessel to maneuver to starboard in head-on situations, and recommend to pass behind an obstacle if the obstacle approaches from the starboard side.
To motivate the algorithm to follow these rules, while being free to ignore the specific maneuvering aspects if required in situations where the other vessel violates the \gls{colregs}, we use the obstacle regions in Figure~\ref{fig:ellipticalColregsObsCostRegions} when calculating this cost.
The regions can be interpreted as follows: the margin region is allowable to enter, the safety region is not desirable to enter, while the collision region should not be entered.
Notice that the algorithm will require a larger clearance in situations where the maneuvering rules in the \gls{colregs} are ignored, e.g. if maneuvering to port in a head-on situation.
See \citet{Eriksen2019} for more details on the $\text{align}(\cdot)$ and $\text{avoid}_\text{m}(\cdot)$ terms.

In this article, we introduce the $\text{avoid}_\text{s}(\cdot)$, $\text{tran}_U(\cdot)$ and $\text{tran}_\chi(\cdot)$ terms.
The $\text{avoid}_\text{s}(\cdot)$ term assigns a cost to avoiding static obstacles, while $\text{tran}_U(\cdot)$ and $\text{tran}_\chi(\cdot)$ are transitional cost terms increasing the robustness to noise.
These terms will be discussed in detail in the following two sections.

\subsection{Static obstacle avoidance}
Static obstacles are modeled using an occupancy grid, which allows for easy representation of obstacles with arbitrary shapes like e.g. land and islands.
In addition, static obstacles are padded with a decaying gradient to introduce some smoothness to the static obstacle avoidance function.
Given an occupancy grid $O(\bs p)\in [0,100]$ where $O(\bs p)=100$ and $O(\bs p) = 0$ represents an occupied and empty cell, respectively, we define the static obstacle term as:
\begin{equation}
	\text{avoid}_\text{s}(\bar{\bs \eta}(t)) = \int_{t_0}^{t_0+T} O(\bar{\bs p}(\gamma)) \mathrm{d}\gamma,
\end{equation}
where $t_0$ denotes the initial time and $\bar{\bs \eta}(t) = \begin{bmatrix} \bar{\bs p}(t)\tr & \bar\psi(t) \end{bmatrix}\tr$.

\subsection{Speed and course transitional costs}
In order to improve the robustness to noise on obstacle estimates, transitional cost is included in the objective function, which penalizes changing the planned trajectory from iteration to iteration.
In \citet{Eriksen2019}, a single transitional cost term is used, which introduces a cost if one selects a different speed and/or course than the one closest to the one selected in the previous iteration.
Note that the trajectory prediction is based on sampling the possible acceleration of the vessel in the current iteration, which implies that the exact trajectory selected in the previous iteration may not exist in the current search space.

Here, it is proposed to split the transitional cost term into separate speed and course terms.
This motivates the algorithm to not alter the course if the speed is changed and vice versa, which would not be the case when using a single transitional cost term.
The transitional cost terms are defined as:
\begin{equation}
  \text{tran}_U(\bs u_d(t)) = \begin{cases}1, & \int_{t_0}^{t_0 + T_1} \left|U_d(\gamma) - U_d^-(\gamma)\right|\mathrm{d}\gamma > e_{U,\min} \\ 0, & \text{else},\end{cases}
\end{equation}
\begin{equation}
  \text{tran}_\chi(\bs u_d(t)) = \begin{cases}1, & \int_{t_0}^{t_0 + T_1} \left|\chi_d(\gamma) - \chi_d^-(\gamma)\right|\mathrm{d}\gamma > e_{\chi,\min}\\ 0, & \text{else},\end{cases}
\end{equation}
with $\bs u_d(t) = \begin{bmatrix} U_d(t) & \chi_d(t) \end{bmatrix}\tr$. The variables $U_d^-(t)$ and $\chi_d^-(t)$ denote the current desired velocity trajectory tracked by the vessel controllers, and $T_1$ is the step time of the first trajectory maneuver. 
The variables $e_{U,\min}$ and $e_{\chi,\min}$ denote the minimum difference between the current desired velocity trajectory and the candidates:
\begin{equation}
  \begin{aligned}
    e_{U,\min} &= \underset{\bs u_d(t) \in \mathcal{U}_d}{\text{min}} \int_{t_0}^{t_0+T_1} \left|U_d(\gamma) - U_d^-(\gamma)\right|\mathrm{d}\gamma \\
    e_{\chi,\min} &= \underset{\bs u_d(t) \in \mathcal{U}_d}{\text{min}} \int_{t_0}^{t_0 + T_1} \left|\chi_d(\gamma) - \chi_d^-(\gamma)\right|\mathrm{d}\gamma.
  \end{aligned}
\end{equation}
	

\section{Experimental results} 
\label{sec:experimental_results}
The modified \gls{bcmpc} algorithm was tested in full-scale experiments in the Trondheimsfjord in Norway on the 27\textsuperscript{th} of September 2018.
This section describes the experimental setup and presets results from the experiments.

\subsection{Experimental setup}
\begin{figure}
	\centering
	\includegraphics[width=\linewidth]{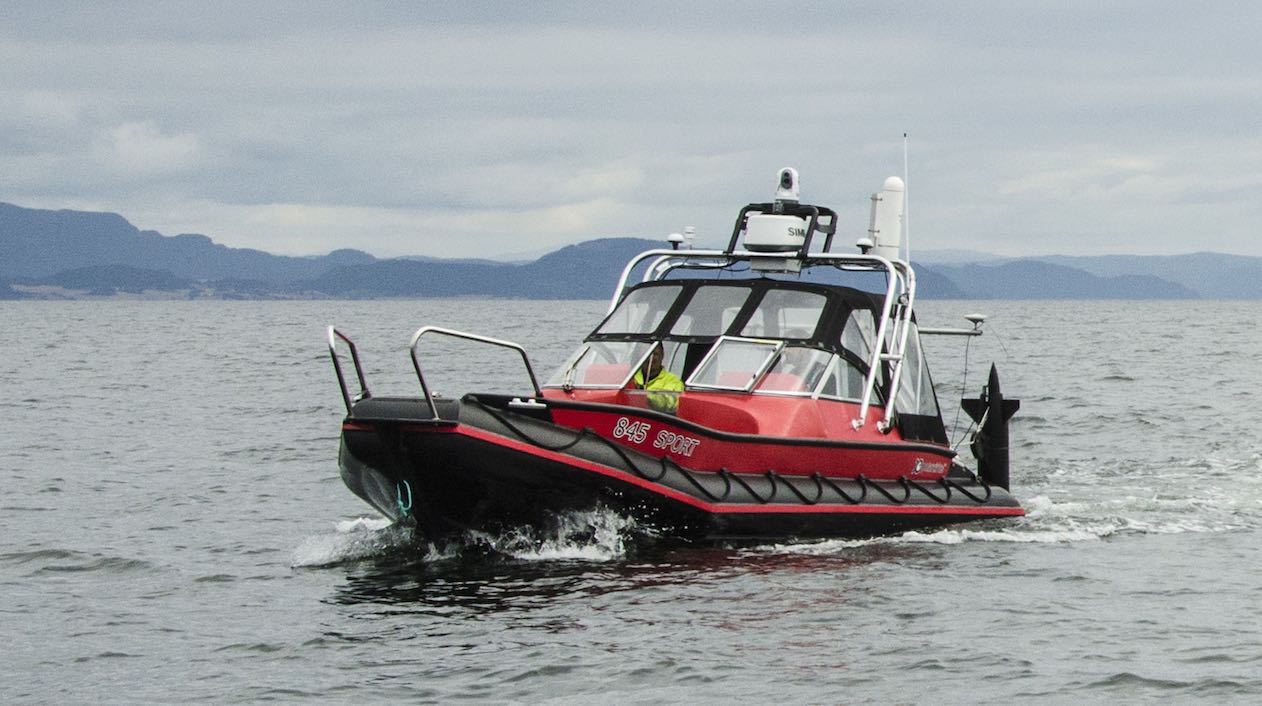}
	\caption{\label{fig:Telemetron}The Telemetron ASV, owned and operated by Maritime Robotics. Courtesy of Maritime Robotics.}
\end{figure}

\begin{figure}[t]
	\centering
	\includegraphics[width=\linewidth]{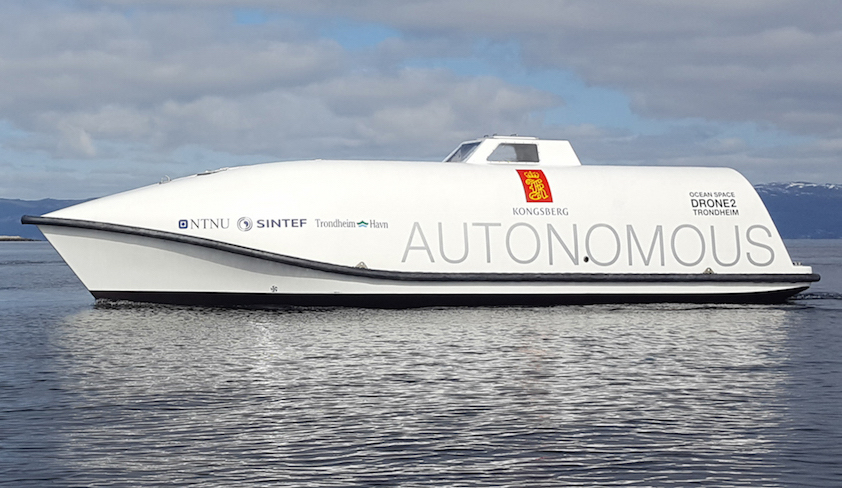}
	\caption{\label{fig:OSD}The Kongsberg Seatex Ocean Space Drone 2, which is identical to the \acrfull{osd1}. Courtesy of Kongsberg Seatex.}
\end{figure}

The experimental setup was similar to the setup reported in \citet{Eriksen2019}, using the Telemetron \gls{asv} from Maritime Robotics as the ownship and the \gls{osd1} from Kongsberg Seatex as the moving obstacle.
In addition, virtual static obstacles, expanded with a padding radius, were used to emulate static obstacles.
The padding radius was selected as $150~\si{\meter}$ in most of the experiments.
Notice that this padding radius only relates to static obstacles and that safety margins for moving obstacles are enforced by the obstacle regions in Figure~\ref{fig:ellipticalColregsObsCostRegions}.
The Telemetron \gls{asv}, shown in Figure~\ref{fig:Telemetron}, is a $26$-foot high-speed \gls{asv} capable of speeds up to $18~\si{\meter\per\second}$ and equipped for both manned and unmanned operations.
The \gls{osd1}, shown in Figure~\ref{fig:OSD}, is a modified offshore lifeboat with a length of $12~\si{\meter}$, and was steered at a constant speed of $5$ knots during the experiments.
The \gls{osd1} played the role of a moving obstacle in the experiments, and was detected and tracked using a radar-based tracking system, which is discussed in detail in~\citet{Wilthil2017} and \citet{wilthil2019}. 
Both the \gls{bcmpc} algorithm and the radar tracking system was implemented using the \gls{ros}, and was run on a processing platform with an Intel® i7 $3.4~\si{\giga\hertz}$ CPU running Ubuntu 16.04 Linux onboard the Telemetron \gls{asv}.
See Table~\ref{tab:HW} for specifications on the Telemetron \gls{asv} and the sensor system.
\begin{table}[t]
\caption{\label{tab:HW}Telemetron ASV specifications.}
{\small
\begin{tabularx}{\linewidth}{lrX}
\toprule
\textbf{Component} & & \textbf{Description} \\
\midrule
Vessel hull                   &   & Polarcirkel Sport 845 \\
  \hspace{1cm}Length              & & $8.45~\si{\meter}$  \\
  \hspace{1cm}Width             & & $2.71~\si{\meter}$  \\
  \hspace{1cm}Weight              & & $1675~\si{\kilogram}$ \\
Propulsion system               &   & Yamaha $225$~HP outboard engine \\
  \hspace{1cm}Motor control           &   & Electro-mechanical actuation of throttle valve \\
  \hspace{1cm}Rudder control          &   & Hydraulic actuation of outboard engine angle with proportional-derivative (PD) feedback control \\
Navigation system                 &   & Kongsberg Seatex Seapath 330+ \\
Radar                                           & & Simrad Broadband 4G™ Radar \\
Processing platform                             & & Intel® i7 $3.4~\si{\giga\hertz}$ CPU, running Ubuntu 16.04 Linux \\
\bottomrule
\end{tabularx}}
\end{table}

The \gls{bcmpc} algorithm was run at a rate of $0.2~\si{\hertz}$ with the parameters in Table~\ref{tab:BCMPC_param}.
\begin{table}[t]
  \centering
  \caption{\label{tab:BCMPC_param}\Gls{bcmpc} algorithm parameters.}
  {\small
  \begin{tabularx}{\linewidth}{llX}
  \toprule
    \textbf{Parameter}        & \textbf{Value}    & \textbf{Description} \\
  \midrule
    B                         & 3     & Trajectory prediction tree depth \\
    $\bs T$                   & $[ 20 , 30 , 30]~\si{\second}$  & Step time lengths\\
    $\bs N_U$                 & $[ 5 , 1 , 1]$                 & Number of \gls{sog} maneuvers  \\
    $\bs N_\chi$              & $[ 5 , 3 , 3]$                 & Number of course maneuvers  \\
    $T_{\text{ramp}}$         & $1~\si{\second}$                                          & Ramp time  \\
    $T_U$                     & $5~\si{\second}$                                          & \Gls{sog} maneuver length  \\
    $T_\chi$                  & $5~\si{\second}$                                          & Course maneuver length  \\
  \cmidrule{1-3}
    $w_{\text{al}}$                  & $1.5$                                                       & Align weight  \\
    $w_{\text{av,m}}$                & $6000$                                                    & Moving obstacle avoid weight  \\
    $w_{\text{av,s}}$                & $30$                                                    & Static obstacle avoid weight  \\
    $w_{\text{t},U}$                 & $2100$                                                    & \gls{sog} transitional cost weight  \\
    $w_{\text{t},\chi}$              & $1050$                                                    & Course transitional cost weight  \\
  \cmidrule{1-3}
    $a_0$               & $50~\si{\meter}$    & Collision region major axis \\
    $a_1$               & $150~\si{\meter}$   & Safety region major axis\\
    $a_2$               & $250~\si{\meter}$   & Margin region major axis\\
    $b_0$               & $25~\si{\meter}$    & Collision region minor axis\\
    $b_1$               & $75~\si{\meter}$    & Safety region minor axis\\
    $b_2$               & $125~\si{\meter}$   & Margin region minor axis\\
    $d_{\text{COLREGs}}$       & $100~\si{\meter}$   & \Gls{colregs} expansion\\
  \bottomrule
  \end{tabularx}}
\end{table}
At sea, vessels typically maneuver with large margins, making it safe to run the \gls{bcmpc} algorithm at this rate.
Furthermore, the sample time of the radar is $2.5~\si{\second}$, which together with the dynamics of the tracking system algorithms results in the closed-loop time delay being dominated by the obstacle detection and tracking system.
With the given tuning parameters, the \gls{bcmpc} algorithm has a runtime of approximately $0.4~\si{\second}$ (including interfacing the radar tracking system), allowing for a higher rate if sensors providing faster updates are available.
The tuning parameters are quite similar to the ones used in the original algorithm, with the exception of the first step time length, which is selected as $20~\si{\second}$ instead of $5~\si{\second}$ in \citet{Eriksen2019}.
With this tuning, the algorithm plans for making one maneuver of $5~\si{\second}$ at the current time and keeping a constant course until $20~\si{\second}$ have passed, rather than planning to do a new maneuver after only $5~\si{\second}$.
This represents a more ``maritime'' way to maneuver compared to performing rapid consecutive maneuvers, and the transitional cost terms will motivate the algorithm to keep a constant course rather than selecting a new planned maneuver.
Notice, however, that the algorithm is still free to choose a new maneuver every $5~\si{\second}$, but the transitional cost terms will favor keeping constant speed and course.
To avoid that the vessel controller limited the performance of the \gls{colav} system, we used a model-based speed and course controller shown to have high performance for high-speed \glspl{asv} \citep{Eriksen2018b}.

During the experiments, we tested four different scenarios:
\begin{enumerate}
	\item A static-only scenario with two static obstacles.
	\item A head-on situation with the \gls{osd1} and four static obstacles.
	\item A crossing situation with the \gls{osd1} and one static obstacle.
	\item An overtaking situation with the \gls{osd1} and one static obstacle.
\end{enumerate}
The desired speed of the Telemetron \gls{asv} was $5~\si{\meter\per\second}$ in all the scenarios, except the overtaking scenario where the desired speed was $8~\si{\meter\per\second}$.

\subsection{Scenario 1}
\begin{figure}[t]
	\centering
	\includegraphics[width=\columnwidth]{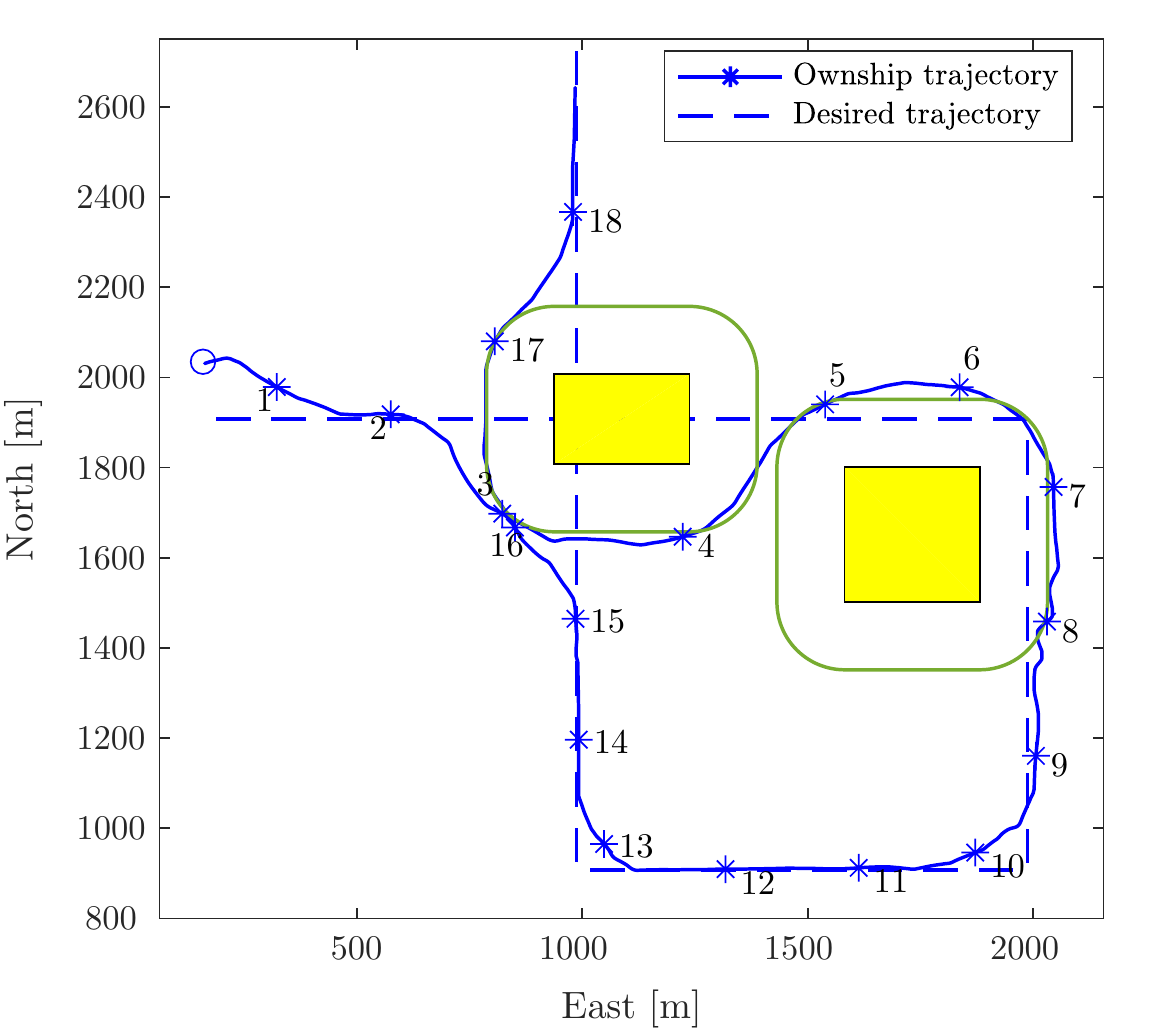}
	\caption{\label{fig:static_only}Scenario 1: Static-only scenario.
The desired trajectory intersects with two obstacles, which the ownship successfully avoids.
The blue circle denotes the initial position, while the text and asterisks mark each $60~\si{\second}$ of the experiment.
The yellow patches show the static obstacles, while the dark green contour lines show the padding regions.}
\end{figure}

Scenario 1 is shown in Figure~\ref{fig:static_only}. 
Here, two static obstacles block the desired trajectory, requiring the \gls{bcmpc} algorithm to circumvent the obstacles.
This scenario may seem a bit unrealistic, since the high-level planner and mid-level \gls{colav} algorithm should plan paths which avoid static obstacles.
However, the \gls{bcmpc} algorithm must be able to avoid static obstacles in order to stay safe in situations where we deviate from the desired trajectory, e.g. when avoiding moving obstacles or in situations where the mid-level algorithm is unable to produce a solution.
The ownship converges to the desired trajectory before avoiding the first static obstacle by maneuvering to starboard.
It would probably have been better to maneuver to port, since this would avoid having to pass through the narrow channel between the first and the second obstacle.
The \gls{bcmpc} algorithm does, however, have a limited planning horizon of $80~\si{\second}$ with the current tuning parameters, which makes it unaware of the narrow channel when making the decision of maneuvering to starboard.
Subsequently, the ownship converges towards the desired trajectory and passes the second obstacle by having a small distance to the desired trajectory, which resides slightly inside the padding region of the static obstacle.
After passing the second obstacle, the ownship converges to the desired trajectory, before avoiding the first obstacle once again.

\subsection{Scenario 2}
\begin{figure}[t]
	\centering
	\includegraphics[width=\columnwidth]{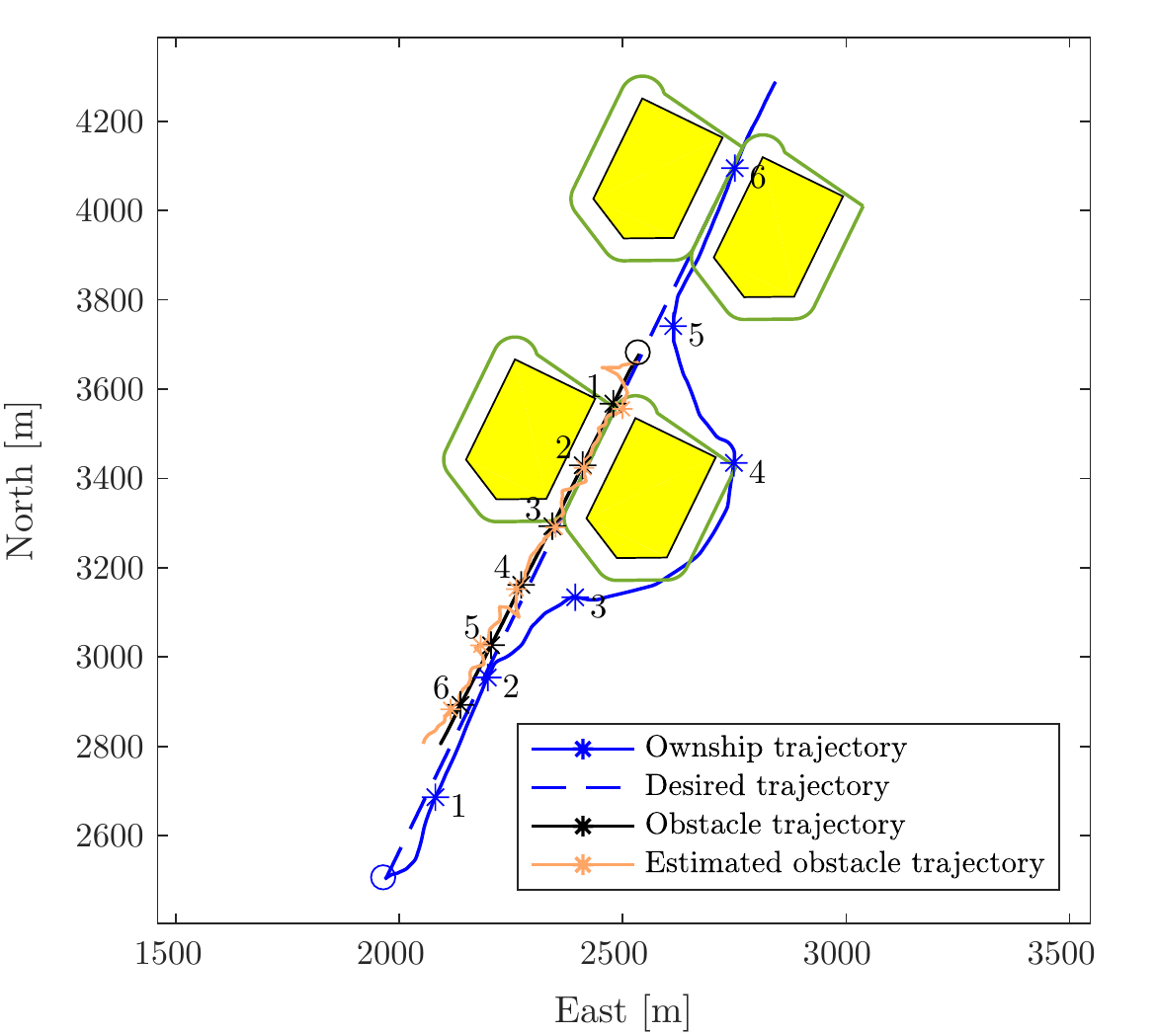}
	\caption{\label{fig:head_on}Scenario 2: Head-on situation.
The desired trajectory passes through a narrow channel, which is blocked by the \gls{osd1}.
The circles denote the initial positions, while the text and asterisks mark each $60~\si{\second}$ of the experiment.
The yellow patches show the static obstacles, while the dark green contour lines show the padding regions.}
\end{figure}

Scenario 2 is a head-on situation where the desired trajectory goes through a narrow channel composed by two static obstacles, and the channel entry is blocked by the \gls{osd1}.
In this scenario, the padding distance was selected as $50~\si{\meter}$ in order to create the narrow channel between the obstacles.
As shown in Figure~\ref{fig:head_on}, the ownship avoids the \gls{osd1} by maneuvering to starboard and hence complying with the \gls{colregs}.
Following this turn, the first static obstacle is passed on the east side.
The ownship returns to the desired trajectory and travels through the channel composed by the two last static obstacles.

\subsection{Scenario 3}
\begin{figure}[t]
	\centering
	\includegraphics[width=\columnwidth]{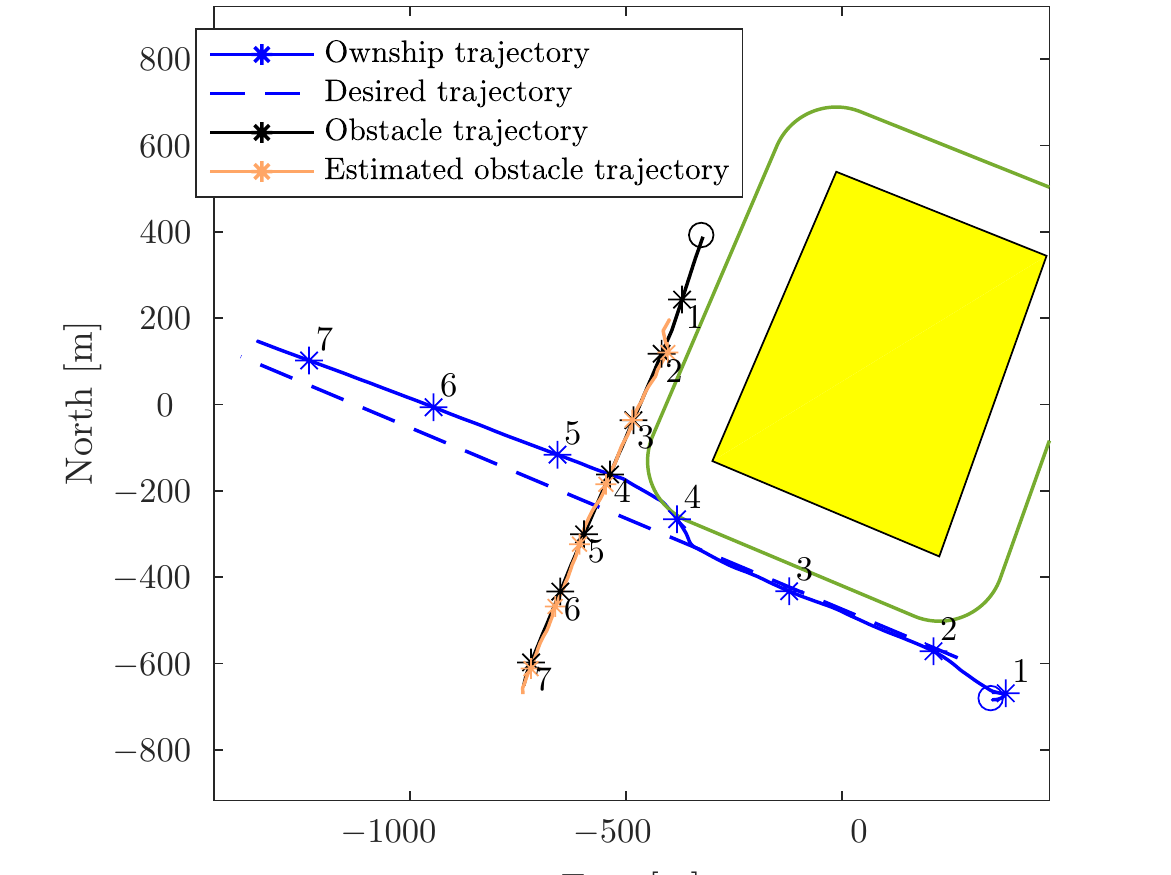}
	\caption{\label{fig:crossing}Scenario 3: Crossing situation.
The desired trajectory intersects with the \gls{osd1}, which approaches from starboard.
The static obstacle encloses Munkholmen, which is a small island located in the Trondheimsfjord.
The circles denote the initial positions, while the text and asterisks mark each $60~\si{\second}$ of the experiment.
The yellow patch shows the static obstacle, while the dark green contour line shows the padding region.}
\end{figure}

Scenario 3, shown in Figure~\ref{fig:crossing}, is a crossing situation where the \gls{osd1} approaches from the ownship's starboard side, requiring the ownship to give way to avoid collision according to the \gls{colregs}.
In addition, there is a static obstacle on the starboard side of the ownship, blocking the ownship from maneuvering to starboard early.
In compliance with the \gls{colregs}, the ownship performs a starboard maneuver in order to pass behind the \gls{osd1}, while passing close to the boundary of the static obstacle.
When the \gls{osd1} has been passed, the ownship slowly converges towards the desired trajectory.
The reason for the slow convergence is that the cost that the transitional cost terms introduces is just too large for the algorithm to change to a trajectory with a faster convergence.
This is sometimes observed, but does not compromise safety and is a subject of tuning the transitional cost weights $w_{\text{t},U}$ and $w_{\text{t},\chi}$.

\subsection{Scenario 4}
\begin{figure}[t]
	\centering
	\includegraphics[width=\columnwidth]{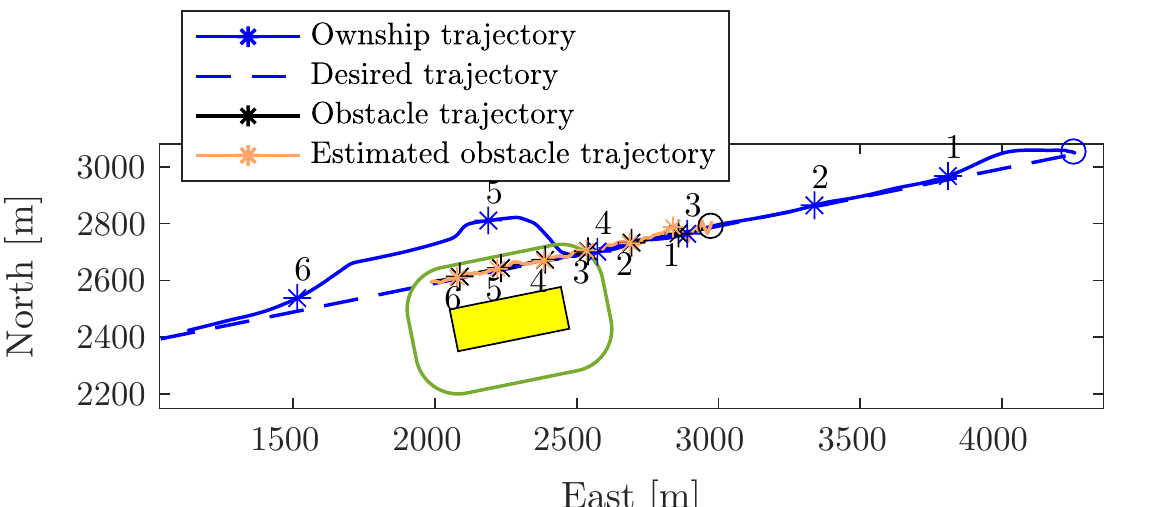}
	\caption{\label{fig:overtaking}Scenario 4: Overtaking situation.
The ownship overtakes the \gls{osd1} by passing on the starboard side, while avoiding the static obstacle.
The circles denote the initial positions, while the text and asterisks mark each $60~\si{\second}$ of the experiment.
The yellow patch shows the static obstacle, while the dark green contour line shows the padding region.}
\end{figure}

Scenario 4 is an overtaking situation where the ownship approaches the \gls{osd1} from behind.
To allow the vessel being overtaken to maneuver to starboard if it finds itself in a separate collision situation, the \gls{bcmpc} algorithm is designed to favor a port turn in overtaking situations.
However, as shown in Figure~\ref{fig:overtaking}, a static obstacle is blocking the port side of the obstacle, which makes the ownship pass the obstacle on its starboard side.
As mentioned, the \gls{bcmpc} algorithm is designed to pass with a larger clearance if passing on the port side rather than the starboard side, which can be seen by comparing this scenario with Experiment 3 in \citep{Eriksen2019}.

\subsection{Experiment summary}
The \gls{bcmpc} algorithm is able to avoid collisions in all the scenarios, while converging to the desired trajectory when it is not obstructed by obstacles.
The resulting ownship trajectories are clear and generally show the intension of the \gls{bcmpc} algorithm.
The ownship trajectories are, however, a bit wobbly when the algorithm traverses alongside static obstacles.
The reason for this is that the trajectory search space consists of a finite number of trajectories, of which none may traverse exactly parallel to the static obstacle.
This results in that the algorithm sometimes choose to ``zig-zag'' along static obstacles, as seen in Scenario 1.
In the usual case where the mid-level algorithm would recompute a collision-free trajectory circumventing the obstacles, the \gls{bcmpc} algorithm would however be able to traverse smoothly along the obstacles by following the desired trajectory.
Also, due to algae growth on the hull, the vessel dynamics had changed quite a bit since the model-based vessel controller was tuned, which also contributed to wobbling in the form of course overshoots.

\begin{table}[t]
\caption{\label{tab:experimentSummary}Minimum distance to obstacles. *The padding distance in Scenario 2 is $50~\si{\meter}$.}
{\small
\begin{tabular}{lll}
\toprule
\textbf{\begin{tabular}{@{}l@{}} Scenario \\ number \end{tabular}} & \textbf{\begin{tabular}{@{}l@{}} Minimum distance \\ to static obstacles \end{tabular}} & \textbf{\begin{tabular}{@{}l@{}} Minimum distance \\ to moving obstacle \end{tabular}} \\
\midrule
Scenario 1 & $130.4~\si{\meter}$ & -- \\
Scenario 2 & $31.3~\si{\meter}$* & $167.1~\si{\meter}$ \\
Scenario 3 & $148.6~\si{\meter}$ & $76.1~\si{\meter}$ \\
Scenario 4 & $115.9~\si{\meter}$ & $145.3~\si{\meter}$ \\
\bottomrule
\end{tabular}}
\end{table}
As seen in Table~\ref{tab:experimentSummary}, the ownship travels inside the padding region of the static obstacles.
This is to be expected, since the objective function is only sensitive to the static obstacles when the trajectory resides inside of the padding region.
Hence, the padding region and static avoidance weight $w_{\text{av,s}}$ should be selected such that a sufficient safety margin is achieved.
A formulation with multiple regions with different gradients, as for moving obstacles, could make it easier to tune the algorithm to obtain a desired safety margin to static obstacles.
The required distance to the moving obstacle is a bit more complex to discuss, since the obstacle regions sizes depend on the relative bearing.
The ownship does, however, stay outside of the safety region in the head-on and crossing scenarios (scenarios 2 and 3), while we slightly enter the safety region in the overtaking scenario (Scenario 4).

\section{Conclusion and further work}\label{sec:Conclusion}
In this article, we have presented two modifications to the \gls{bcmpc} algorithm for \gls{asv} \gls{colav}.
The first modification allows the algorithm to avoid static obstacles in the form of an occupancy grid.
The second modification concerns improved transitional cost terms by introducing transitional cost in speed and course separately, motivating the algorithm to not change the course if the speed is changed and vice versa.
In addition, the algorithm tuning has been changed in order to obtain more ``maritime'' maneuvers and better utilize the transitional cost terms.
The modified \gls{bcmpc} algorithm is tested in full-scale experiments in the Trondheimsfjord in Norway.
A moving obstacle is detected and tracked using a radar-based system, while virtual static obstacles are added in the \gls{colav} system.
Four different scenarios were tested in experiments, all of which provided good results.

In \citet{Eriksen2019c}, the authors have used the \gls{bcmpc} algorithm described in this article in a hybrid architecture, demonstrating \gls{colav} compliant with \gls{colregs} rules 8 and 13--17 in simulations.
In the future, we would like to perform an extensive simulation study of the \gls{bcmpc} algorithm, in order to analyze the algorithm's performance in greater detail.

\section*{Acknowledgments}
This work was supported by the Research Council of Norway through project number 244116 and the Centres of Excellence funding scheme with project number 223254. The authors would like to express great gratitude to Kongsberg Seatex and Maritime Robotics for providing high-grade navigation technology, the Telemetron \gls{asv} and the \gls{osd1} at our disposal for the experiments.

\bibliographystyle{mic}
\bibliography{references}

\end{document}